\documentclass[aps,twocolumn,english,superscriptaddress,citeautoscript,showkeys,preprintnumbers,amsmath,amssymb,floatfix,footinbib,prb]{revtex4-2}
\usepackage{graphicx}
\usepackage{xcolor}
\usepackage{soul}
\usepackage{amssymb,amsmath}
\usepackage{hyperref}
\hypersetup{colorlinks=true,linkcolor=red,citecolor=blue}
\usepackage[utf8]{inputenc}
\usepackage[english]{babel}
\usepackage{txfonts}
\usepackage{multirow}
\usepackage{orcidlink}

\DeclareUnicodeCharacter{0303}{HERE!HERE!}

\begin{document}

\title{First-principles modeling of ordering in Fe-rich Fe-Si alloys}

\author{D. M. Khodachenko\,\orcidlink{0009-0001-5502-7687}}
\affiliation{Christian Doppler Laboratory for digital material design guidelines for mitigation of alloy embrittlement, Materials Center Leoben Forschung GmbH, 8700, Leoben, Austria}
\author{A.~V. Ruban\,\orcidlink{0000-0002-3880-0965}}
\affiliation{Materials Center Leoben Forschung GmbH, 8700, Leoben, Austria}
\affiliation{KTH Royal Institute of Technology, 10044, Stockholm, Sweden}
\author{F. Moitzi\,\orcidlink{0000-0003-0558-7966}}
\affiliation{Christian Doppler Laboratory for digital material design guidelines for mitigation of alloy embrittlement, Materials Center Leoben Forschung GmbH, 8700, Leoben, Austria}
\author{O. E. Peil\,\orcidlink{0000-0001-9828-4483}}
\affiliation{Materials Center Leoben Forschung GmbH, 8700, Leoben, Austria}
\author{V. I. Razumovskiy\,\orcidlink{0000-0002-2157-3110}}
\affiliation{Christian Doppler Laboratory for digital material design guidelines for mitigation of alloy embrittlement, Materials Center Leoben Forschung GmbH, 8700, Leoben, Austria}
\begin{abstract}

The atomic ordering in Fe-rich Fe-Si alloys strongly influences 
their magnetic properties and mechanical behavior. While general features of long-range atomic ordering are known from
the experimentally established phase diagram in this composition range, detailed atomic configurations and driving 
forces behind them remain speculative. Here, we address this gap by using 
effective cluster interactions (ECI) obtained from density functional theory (DFT) calculations and subsequent statistical
thermodynamics modeling to unravel the interplay between magnetism, composition, and temperature in body-centered cubic
(bcc) Fe--Si alloys up to 20~at\% Si.  Our analysis reveals that the ordering behavior is governed by a strong coupling
between magnetic order, lattice expansion, and Si concentration, with specific ECI shells switching sign or magnitude as
the magnetic state changes, which is reflected in the phase equilibria and
ordering behavior of Fe-Si alloys. In particular, we demonstrate that D0$_3$ ordering is,
in fact, primary in the totally ordered ferromagnetic state, while the B2 type of ordering is promoted in
the paramagnetic state at elevated temperatures.  The temperature-dependent evolution of these ordering patterns, 
obtained from Monte Carlo simulations based on the magnetically resolved ECIs, reproduces the experimentally observed
phase equilibria and short-range order characteristics. By explicitly linking the magnetic state with the effective interactions and 
macroscopic phase behavior, this work establishes a predictive framework for an \textit{ab initio}-based description of ordering 
in Fe-rich Fe--Si alloys and highlights the decisive role of magnetism in ordering phenomena in this system.

\end{abstract}

\date{\today}


\maketitle

\section{Introduction}

The unique electromagnetic properties of Fe-rich Fe-Si alloys make them a perfect choice for the production of transformers and electric motors. These so-called electrical steels have improved energy efficiency due to reduced core losses and enhanced magnetic permeability.
These are bcc random alloys with a Si content usually below 12~at\%. At the same time, a large body of experimental data suggests
that the origin of these magnetic properties is related to a specific B2-type atomic short-range order (SRO), i.e., the existence
of pairs of Si atoms at the second coordination shell of the bcc lattice (see, for instance, \cite{ershov2006,ershov2012}). If Si content exceeds 6~wt\% or 12~at\%, these alloys suffer from significantly reduced ductility, preventing the development of steels with more advanced properties. The origin of brittleness in such high-Si electrical steels is also attributed to atomic ordering, namely, a precipitation of the Fe$_3$Si-D0$_3$ ordered phase, which, according to the existing phase diagrams should occur at temperatures below around $800$~K at this concentration \cite{ohnuma2012,han2022,FeSi_calphad}. This means that knowing the details of the atomic ordering phenomena in  Fe-rich Fe-Si alloys and understanding their origin is of prime importance for the scientifically guided development of electrical steels.

Several experimental investigations of atomic SRO in random Fe–Si alloys with compositions corresponding to electrical steels
have been reported (see, in particular, \cite{exp_sro,ershov2012}). Although detailed information on the atomic SRO
is limited due to experimental restrictions for dilute alloys, the prevailing consensus is that a B2-type SRO dominates
in alloys with Si content below 8~at\%, whereas a D0$_3$-type SRO emerges at higher Si concentrations.
At the same time, theoretical studies of atomic ordering in these alloys remain scarce, likely due to the complex
nature of interatomic interactions driven by magnetism, which is extremely challenging to model accurately
in \textit{ab initio} calculations. In fact, there exists only one \textit{ab initio} investigation of effective interactions and atomic SRO in Fe–(4–10)~at\% Si by Gorbatov \textit{et al.} \cite{gorbatov2011}. As demonstrated in that work,
increasing magnetic order promotes D0$_3$-type atomic ordering; that is, only the B2-type SRO exists in
the paramagnetic state, while upon lowering the temperature, the D0$_3$-type SRO appears in the ferromagnetic state,
consistent with experimental observations.

In this paper, we extend the previous theoretical study to a broader concentration range, up to 20~at\% Si, while also
accounting for the effects of concentration and temperature on the effective interactions, which were neglected
in \cite{gorbatov2011}. This enables us to estimate the boundaries of the ordered phases present in the Fe–Si
phase diagram using subsequent Monte Carlo statistical simulations. The temperature dependence of the ECI is governed
primarily by the temperature-dependent magnetic state of the alloys and, to a lesser extent, by the corresponding
equilibrium lattice constant.

\section{Methodology}
\label{ch:Method}

In this study, we have used two different but complementary \textit{ab initio} techniques:
(i) The exact muffin-tin orbital (EMTO) method \cite{andersen1994,emto1,emto2,vitos_book_emto}, which
is a multiple scattering technique that enables efficient calculations of electronic
structure and energetics of chemically and magnetically random alloys within the coherent potential approximation (CPA) \cite{cpa1,cpa2,emto3,Zabloudil_book}
as well as ECI \cite{andrei_mohammad,Summary_Ruban}. Here, previously developed workflows \cite{adamant1, adamant2} have been extended in order to automate and streamline such ECI calculations. (ii) The projector augmented wave (PAW) 
method \cite{PAW}, as implemented in the Vienna \textit{ab initio} simulation
package (VASP) \cite{kresse1999,kresse1994,kresse1996}, which provides the accurate energetics of model systems restricted practically only by the usual approximations of density functional theory (DFT) \cite{HohenbergKohn,KohnSham}.

\subsection{Multiple scattering theory calculations by EMTO}
\label{ch:EMTO}

Three types of calculations have been performed using the EMTO method: 1) Total energies of random Fe-Si alloys within the CPA, 2) total energies within the locally self-consistent Green's function (ELSGF) technique \cite{ELSGF,LSGF_main,LSGF1}, and 3) effective cluster calculations within the screened generalized perturbation method (SGPM) \cite{GPM_main1,GPM_main2,ruban2004}. The screening constants for EMTO-CPA and for effective pair interactions (EPI) have been computed using ELSGF.

To treat the entire range of magnetizations $m$, between the paramagnetic (PM, $m = 0$) and saturated ferromagnetic (FM, $m = 1$), we have
used the alloy analogy for Fe atoms in the form Fe$_{1-y}^{\uparrow}$Fe$_{y}^{\downarrow}$, with $m=1-2y$.
The case $m=0$ and $y=0.5$ corresponds to the disordered local moment (DLM) model of the paramagnetic state.
For intermediate values $0 < m < 1$, the magnetic configuration is treated using partial DLM (PDLM). 
Let us note that upon approaching saturation, $m \rightarrow 1$, the DFT-DLM self-consistent calculations become
unstable, resulting in the tendency to converge to the FM ground state due to a sign change in the minority magnetic moment.
In such cases, the minority magnetic moment has been fixed to its DLM magnitude.
We have used both the local density approximation (LDA) \cite{LDA} and the generalized gradient approximation by Perdew, Burke and Ernzerhof (GGA-PBE) \cite{PBE} exchange-correlation (XC) potentials to estimate
the error related to the XC functional. 

\subsection{Configurational Hamiltonian and effective cluster interactions}
\label{ch:GPM}

The statistical thermodynamic simulations of chemical ordering in alloys are based on the
Ising-type configurational Hamiltonian:
\begin{align}
    \nonumber
    \mathcal{H}_{\text{conf}} = &\frac{1}{2}\sum_{p} V_p^{(2)} \sum_{i,j\in p}\delta c_i\delta c_j\\ +&\frac{1}{3}\sum_t\ V_t^{(3)}\sum_{i,j,k\in t} \delta c_i\delta c_j\delta c_k+... \; ,
\end{align}
where the alloy configuration is given in terms of concentration fluctuations $\delta c_i=c_i-c$. Here, $c$ is the solute concentration, and $c_i$ is the occupation number of site $i$, which takes a value of one when the site is occupied by a given atom type, and zero otherwise.
The effective cluster interactions (ECI) $V^{(n)}_f$
are associated with the configurational energy difference of an $n$-site cluster of type $f$. The interactions have been determined in SGPM calculations based on the EMTO-CPA method \cite{ruban2004,andrei_mohammad} as well as in direct total energy supercell calculations within VASP.

In this work, we account for two contributions to ECI: chemical and strain-induced. The chemical interactions have been computed on the ideal undistorted bcc lattice, and are therefore related only to the energy contributions of atomic rearrangements. The remaining strain-induced part
of the effective interactions has been obtained in VASP supercell total energy calculations, as described below.

Effective pair interactions (EPI) contain an additional contribution from the inter-site screened Coulomb 
interactions:
\begin{equation}
    V_{p}^{(2)-\text{scr}} = \alpha_{p}^\text{scr} \frac{e^2 \Delta {q}^2}{S} ,
\end{equation}
where $\Delta q$ is the alloy charge transfer and $S$ is the Wigner-Seitz radius. The screening constants, $\alpha_{p}^\text{scr}$,
have been determined for the dilute alloy composition as well as for 12~at\% of Si in the FM and PM states in the corresponding ELSGF calculations \cite{Scr1,Scr2}.

\subsection{ECI by VASP supercell calculations}
\label{ch:VASP}

PAW calculations within VASP have been employed to obtain the effective chemical and strain-induced
interactions in the dilute limit of Si in Fe-Si alloys. All PAW calculations have been
performed using the GGA-PBE XC functional \cite{PBE}.

In general, ECIs of order $n$ for a given cluster $f$ can be obtained in
the dilute limit of Si in Fe from the total energies of supercell calculations as
\begin{eqnarray}
      V^{(n)}_f &=& (-1)^n E^{\text{Fe}}  \\ \nonumber
      &+&\sum_{k=1}^n (-1)^{n-k} \sum_{1 \leq j_1 < ... < j_k \leq n} E^{\text{Si}_{j_1}...\text{Si}_{j_k}[\text{Fe}]}_f ,  
\end{eqnarray}
where $E^{\text{Si}_{j_1}...\text{Si}_{j_k}[\text{Fe}]}_f$ is the total energy of
$k$ Si atoms embedded in the Fe supercell and occupying positions $j_1$, $j_2$, etc. in the cluster $f$.

So, the EPI in this case are given by
\begin{equation} \label{eq:V_sc}
V^{(2)}_p = E^{\rm SiSi[Fe]}_p -2 E^{\rm Si[Fe]} + E^{\rm Fe},
\end{equation}
where $E^{\rm SiSi [Fe]}_p$ is the total energy of an Fe supercell with two Si atoms at the $p$-th
coordination shell; $E^{\rm Si[Fe]}$ is the total energy of an Fe supercell with one Si atom, and
$E^{\rm Fe}$ the total energy of Fe. If the total energies are obtained for an unrelaxed ideal
underlying lattice, $V^{(2)}_p$ represents the chemical interaction, which can be compared
with the corresponding SGPM one. If the structures in calculations are \textit{locally} 
relaxed, $V^{(2)}_p$ is the total interaction, which is the sum of the chemical and strain-induced
interactions. In a similar way, one can determine multi-site interactions.

In this work, a $432$-atom $6 \times 6 \times 6 (\times 2)$ supercell has been used to calculate EPI on 
the underlying bcc lattice in the FM state. The calculations have been done for the 
lattice constant of $2.85$~\AA. The $k$-point grid for the Brillouin zone integration has been set to $4 \times 4 \times 4$.

\begin{table*}
\caption[Table I]{Chemical EPI in Fe-Si alloys at dilute Si concentration obtained from supercell (SC) and SGPM calculations, given in mRy. In the case of SGPM, both LDA and GGA exchange-correlation potentials have been tested. The SC results for the paramagnetic state are obtained for DAFM structures.}
\label{tab:EPI}
\begin{ruledtabular}
\begin{tabular}{|c|c|c|c|c|c|c|c|c|c|}
\hline
& \multicolumn{3}{c|}{ferromagnetic state}
& \multicolumn{3}{c|}{paramagnetic (DLM) state}
& \multicolumn{3}{c|}{nonmagnetic state} \\
\hline
Shell
& SC-GGA & SGPM-LDA & SGPM-GGA
& SC-GGA & SGPM-LDA & SGPM-GGA
& SC-GGA & SGPM-LDA & SGPM-GGA \\
\hline
1 (111)
& 31.26 & 26.45 & 28.87
& 45.92 & 37.58 & 35.69
& 45.00 & 45.96 & 45.28 \\
2 (200)
& 16.51 & 13.60 & 14.49
&  6.25 & 7.25  &  8.10
&  0.84 & -0.24 & -0.06 \\
3 (220)
& -0.46 & -0.82 & -0.06
& -0.43 & -0.45 & -0.26
&  1.23 &  0.55 &  0.73 \\
4 (311)
& -0.88 & -0.72 &  0.87
&  1.49 &  0.37 &  0.50
& -0.04 &  0.10 &  0.22 \\
5 (222) 
&  0.77 & -0.06 & -1.07
& -0.12 &  0.69 &  0.94
& -0.73 & -0.04 &  0.94 \\
\hline
\end{tabular}
\end{ruledtabular}
\end{table*}

To estimate effective interactions in the PM state, we have performed total energy calculations for $256$-atom $4 \times 4 \times 4 (\times 2)$ supercells with double antiferromagnetic (DAFM) layering in the [$001$] direction.  In this structure, the nearest-neighbor
spin-spin correlation function is equal to zero.  Although the spin-spin
correlation functions are quite large for the other coordination shells in the
DAFM structure, the corresponding magnetic exchange interactions in the DLM
state are relatively small compared to the nearest-neighbor one \cite{ruban2004}.
This is the simplest supercell model that can be used for a semi-quantitative
test of the ECI in the PM state for the bcc structure.

In the DAFM case, the effective interactions have been obtained as
\begin{equation} \label{eq:V_sc_dafm}
V^{(2)}_p = E^{\rm SiSi [Fe]}_p - E^{\rm Si;Si [Fe]},
\end{equation}
where $E^{\rm Si;Si [Fe]}$ is the total energy of the Fe supercell with the two most distant Si atoms.
In both supercells, the choice of the Si sites has been made in a way to keep the total
number of Fe$^\uparrow$ and Fe$^\downarrow$ equal, keeping the effect on the original DAFM magnetic structure,
and in particular its zero total magnetic moment, as small as possible.

\subsection{Monte Carlo simulations}
\label{ch:MC}
In this work, two different types of Monte Carlo (MC) \cite{metropolis_alg,MC_book_ch,MC_with_SGPM,MC_book_Kawasaki,SGPM_kawasaki_applic} methods have been used to obtain atomic configurations of Fe-Si alloys at finite temperatures:
(i) a conventional Metropolis algorithm \cite{metropolis_alg} has been used to obtain thermodynamic equilibrium configurations, and (ii) a simplified kinetic-MC (kMC) method
based on Kawasaki dynamics \cite{MC_book_Kawasaki,SGPM_kawasaki_applic} has been used to simulate the time evolution. In the latter approach, nearest-neighbor exchange of atoms is employed to represent the effect of atomic diffusion in substitutional alloys.

There are two ways to connect MC steps in such Kawasaki dynamics simulations to real timescales. The first approach utilizes the impurity diffusion coefficient $D_i$ of the solute atom $i$ in an alloy, e.g., in our case the Si impurity diffusion in Fe. With this coefficient, one can determine the diffusion time $\tau$ of an impurity atom using the relation
\begin{equation}
    D_i = \frac{d^2}{\tau}.
\end{equation}
Considering that an MC step corresponds to one nearest-neighbor exchange, the distance $d$ is the real distance of the first coordination shell, which for bcc is $\frac{\sqrt{3}}{2}a$.

Alternatively, MC steps can be connected to the Zener relaxation time  measured in internal friction
experiments \cite{FeAl_FeSi_diff, Diff_FeAl, Diff_FeSi, activE_all}, since the Zener relaxation time and diffusion time are
equivalent (up to a geometric factor $\gamma$ \cite{Diff_FeSi}, depending on the experimental setup) and
both follow the Arrhenius law
\begin{equation}
    \label{eq:Arrh_law}
    \tau = \tau_0 \,\, \text{exp}\left( \frac{Q}{RT}\right).
\end{equation}

Using this approach, one only requires the pre-exponential factor $\tau_0$ and the activation energy $Q$ from experiment in order to estimate the diffusion time for any temperature. In the following Section~\ref{ch:kMC_results}, we use this exact approach in order to estimate the timescales of long-range order (LRO) domain formation.

\section{Results I: Effective interactions and atomic ordering in F\lowercase{e}-S\lowercase{i} }
\label{ch:ResultsI}

\subsection{SGPM vs supercell EPI: FM vs DLM}
\label{ch:EPI_fm_pm_results}

In Table~\ref{tab:EPI}, we compare EPI in the dilute alloy of Si in Fe in the FM, the PM (DLM) and the nonmagnetic states obtained
from the EMTO-CPA SGPM and from PAW supercell (SC-GGA) calculations for a lattice constant of $2.85$~\AA.
In the case of SGPM, we used two different exchange-correlation potentials, namely LDA and GGA-PBE.

The agreement for the strongest EPI at the first two coordination shells between supercell (SC) and SGPM results is fairly good despite the fact that these methods employ totally different \textit{ab initio} techniques. In particular, the effective interactions obtained in SC calculations contain renormalization effects of the local environment of different Si configurations, which are often found to be rather large in alloys containing a combination of magnetic and nonmagnetic elements. Such effects are not contained in SGPM, due to the single-site approximation.

\begin{figure}[t!]
    \centering
    \includegraphics[width=0.45\textwidth]{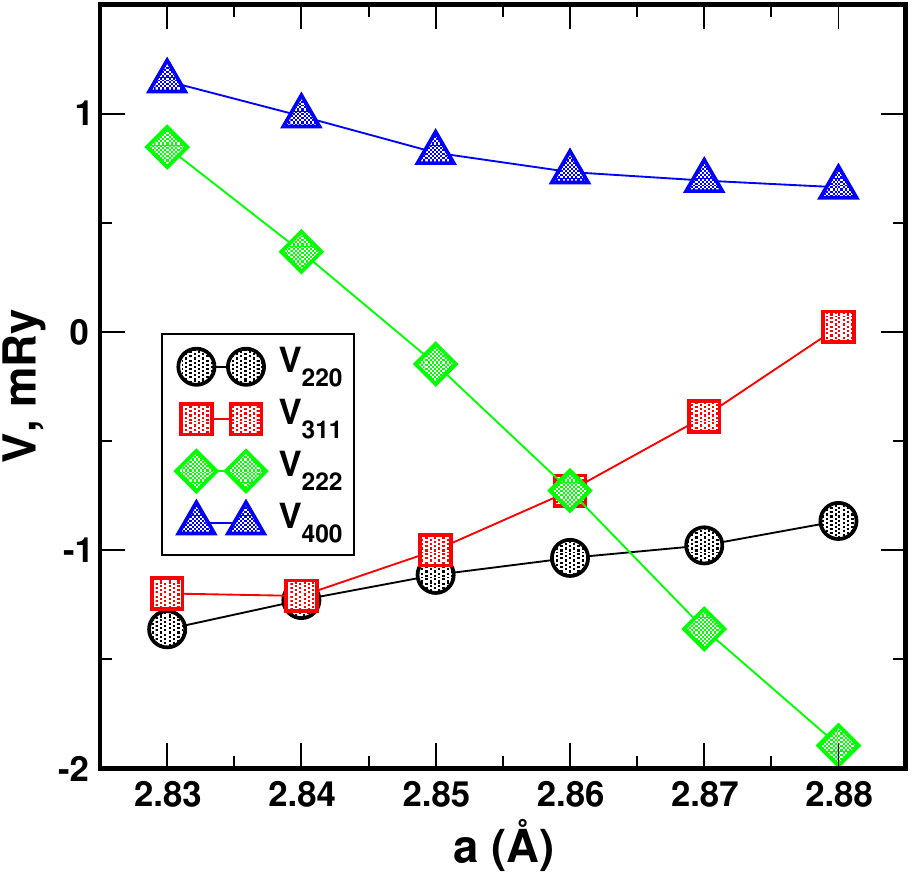}
    \caption{Chemical EPI from SGPM at the third ($220$), fourth ($311$), fifth ($222$), and sixth ($400$) coordination shells for the Fe$_{0.99}$Si$_{0.01}$ alloy in the FM state.}
    \label{fig:v_a}
\end{figure}

However, for more distant EPI, e.g., the fourth and fifth coordination shells in FM, results for GGA-PBE and LDA differ rather significantly, less pronounced in PM and nonmagnetic simulations. One of the reasons for this is the well-known difference between magnetic moments obtained from LDA and GGA-PBE for Fe in the FM state, which contributes to the difference in magnetic exchange interactions \cite{ruban2004}. Such magnetic exchange interactions are implicitly contained in the chemical EPI, therefore, a change in the XC potential evidently also changes the chemical EPI in FM Fe-rich alloys. Note that other choices of GGA functionals, for instance, PBEsol \cite{PBEsol}, may produce other results for the EPI than PBE, as GGA itself is not uniquely defined.

Furthermore, distant EPIs in the FM state are quite sensitive to the lattice constant
in the dilute limit of Si in Fe, as one can see in Fig.~\ref{fig:v_a}. For this reason, we have performed a dedicated study on the impact of this lattice constant dependence on chemical ordering in the following Sections~\ref{ch:grid}, \ref{ch:EPI_grid_results} and \ref{ch:MC_Results}.

The lattice constant dependence of the EPIs is especially pronounced for the $V_{222}$ interaction
(the fifth coordination shell), which is again related to the dependence of EPI on magnetic moments in FM Fe-rich alloys.

In the following sections, we will use the LDA in SGPM calculations, as for Fe it is known to match the experimental low-temperature magnetic moments quite well \cite{Fe_magmom}, while most other potentials, including GGA-PBE, are found to overestimate magnetic moments. Note that the LDA potential also tends to underestimate the equilibrium volumes obtained from \textit{ab initio} simulations, but since we generally use experimental lattice constants in this study this drawback is not an issue.

\begin{figure}[t!]
    \centering
    \includegraphics[width=0.45\textwidth]{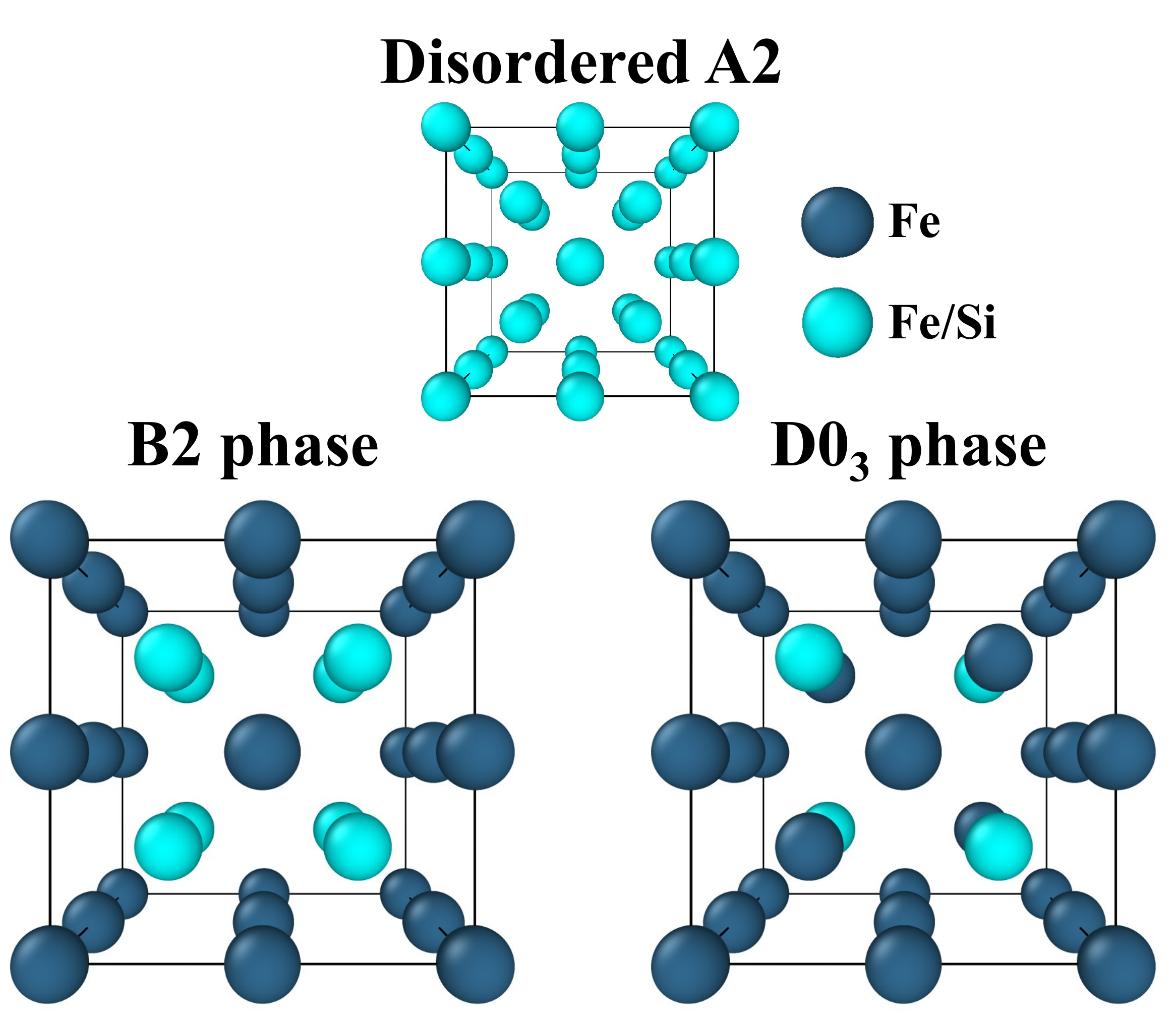}
    \caption{Relevant chemical ordering types in the Fe-rich region of Fe-Si alloys.}
 \label{fig:B2_DO3}
\end{figure}

\subsection{Magnetic state and ordering in Fe-rich Fe-Si alloys: A representative example}
\label{ch:generic_picture}

\begin{table*}
\caption[Table II]{Warren-Cowley SRO parameters $\alpha_i$ at the first eight coordination shells for the totally ordered B2 and D0$_3$ structures. Coordination shells with different signs of SRO parameters for B2 and D0$_3$ phases are marked in bold.}
\label{tab:SRO}
\begin{ruledtabular}
\begin{tabular}{|r|c|c|c|c|c|c|c|c|}
 \hline
   Shell & 1 (111) &  2 (200) & 3 (220) & 4 (311) & 5 (222) & 6 (400) & 7 (331) & 8 (420) \\
 \hline
 B2 &  -1  &  \textbf{1}  &  1  & -1  &  \textbf{1}  &  1  & -1  &  \textbf{1} \\
 \hline
 D0$_3$ & -1/3 & \textbf{-1/3} &  1  & -1/3 & \textbf{-1/3} &  1  & -1/3 &  \textbf{-1/3} \\
 \hline
\end{tabular}
\end{ruledtabular}
\end{table*}

In order to demonstrate the complexity of chemical ordering and the effect of the magnetic state on phase stability in Fe-Si alloys, 
we present the results of two MC simulations for Fe$_{0.88}$Si$_{0.12}$, with ECIs obtained at the corresponding alloy composition
and a fixed lattice constant of $2.87$~\AA{} in the FM and PM (DLM) states. Here, as well as in later sections for temperature- and concentration-dependent ECI, it is found that the FM state promotes D0$_3$-type atomic order, while the PM state promotes B2 ordering (see Fig.~\ref{fig:B2_DO3} for their geometric arrangements). The B2 phase is characterized by two simple cubic sublattices in the bcc superlattice, with one sublattice containing exclusively Fe atoms and the other containing both, while in the D0$_3$ phase,
the bcc lattice is divided into three symmetry-inequivalent sublattices: one simple cubic Fe-sublattice, one fcc-like Fe-sublattice, and one fcc-like Si-sublattice.

\begin{figure}[t!]
    \centering
    \includegraphics[width=0.45\textwidth]{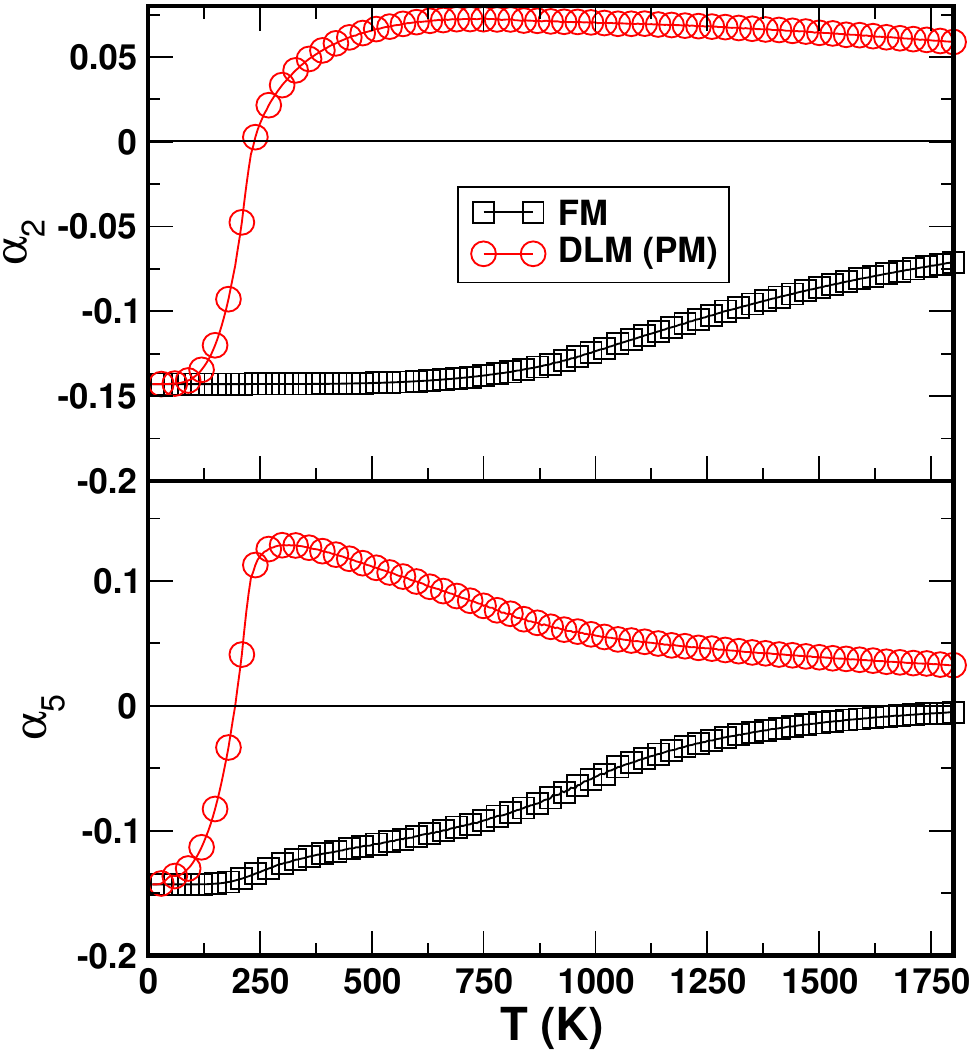}
    \caption{Warren-Cowley SRO parameters of the Fe$_{0.88}$Si$_{0.12}$ alloy at the second and fifth coordination shells obtained in 
    MC simulations with fixed ECI in the FM and DLM states, at a lattice parameter of $2.87$~\AA.}
    \label{fig:a25}
\end{figure}

A given long-range ordered state is characterized by the corresponding LRO parameters. In the B2 phase, the LRO parameter is related to the superstructure vector ($100$) and can be defined as the difference between the concentrations of the majority element Fe on the Si- and Fe-sublattices: 
\begin{equation}
    \eta_{100} = c^{\rm Fe}_{\rm Fe} - c^{\rm Si}_{\rm Fe},
\end{equation}
where, $c^{\alpha}_{\rm Fe}$ is the concentration of Fe on a sublattice $\alpha$. In the case of the D0$_3$ structure, two LRO parameters are required to describe the ordered alloy configuration. The first one is $\eta_{100}$ as in the case of the B2 phase while the other one is related to the ($\frac{1}{2}\frac{1}{2}\frac{1}{2}$) superstructure vector. It can be determined in a similar way, in this case using three different sublattices: Fe-1, Fe-2, and Si. The LRO parameters in the D0$_3$ phase (see Fig.~\ref{fig:B2_DO3}) can then be written as follows:
\begin{eqnarray}
    c^{\rm Fe-1}_{\rm Fe} &=& c_{\rm Fe} + \frac{1}{4} \eta_{100} \\ \nonumber
    c^{\rm Fe-2}_{\rm Fe} &=& c_{\rm Fe} - \frac{1}{4} \eta_{100} + \frac{1}{2}\eta_{\frac{1}{2}\frac{1}{2}\frac{1}{2}} \\ \nonumber
    c^{\rm Si}_{\rm Fe} &=& c_{\rm Fe} - \frac{1}{4} \eta_{100} - \frac{1}{2}\eta_{\frac{1}{2}\frac{1}{2}\frac{1}{2}} .
\end{eqnarray}
Here, $c_{\rm Fe}$ is the total concentration of Fe in the alloy, and $\eta_{\frac{1}{2}\frac{1}{2}\frac{1}{2}}$ the second LRO parameter of the D0$_3$ phase.

In practice, calculations of LRO parameters in MC simulations are not straightforward, as the sublattices are well-defined
only relative to each other. An alternative and convenient way to identify these ordered phases is to use the Warren-Cowley
short-range order (SRO) parameters \cite{cowleySROparam,SRO_fcc_bcc}. In Table~\ref{tab:SRO}, we show the Warren-Cowley SRO parameters for the first eight coordination
shells for totally ordered B2 and D0$_3$ structures.
Although the absolute values of the SRO parameters can be different for arbitrary compositions and ordering states, their sign is fixed.
Thus, the sign of the SRO parameters can be
used to distinguish the phases. In particular, as one can see in Table~\ref{tab:SRO}, the SRO parameters at the second, fifth and eighth coordination shells have different signs between the B2 and D0$_3$ phases, and thus they can be used to identify the type of ordering in statistical MC simulations of thermodynamics.

\begin{figure}[t!]
    \centering
    \includegraphics[width=0.45\textwidth]{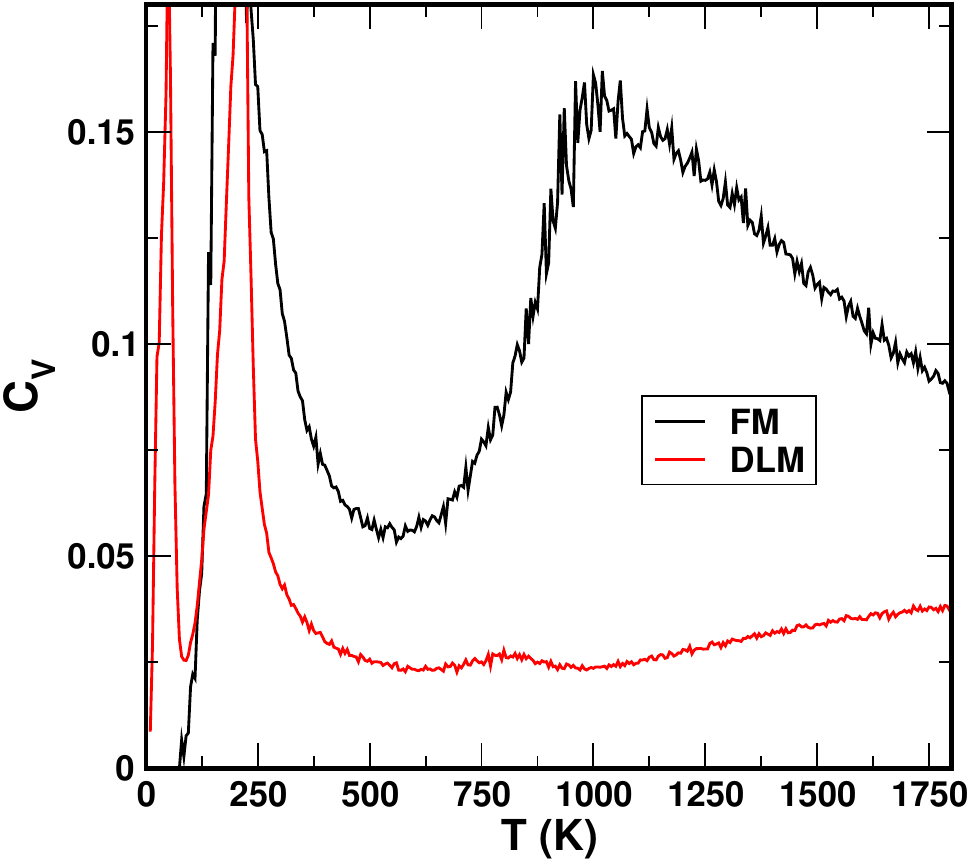}
    \caption{Heat capacity of the Fe$_{0.88}$Si$_{0.12}$ alloy in the FM and DLM states from MC simulations with fixed ECI, at a lattice parameter of $2.87$~\AA.}
    \label{fig:cv_si12}
\end{figure}

In Fig.~\ref{fig:a25}, we show the results for SRO parameters at the second and fifth coordination shells obtained in MC simulations
with FM and DLM ECI. As can be seen, $\alpha_2$ and $\alpha_5$ remain negative for all temperatures in the FM state,
which according to Table~\ref{tab:SRO} corresponds to the D0$_3$-phase. The actual A2-D0$_3$ phase transition can be identified
from the temperature dependence of the SRO parameters as an inflection point. However, it can be obtained with more precision
from the maximum of the corresponding heat capacity $C_V$ displayed in Fig.~\ref{fig:cv_si12}. 
The high-temperature peak of $C_V$ in the FM-state simulation indicates that the A2-D0$_3$ transition occurs at about $1000$~K. Interestingly, this is a primary phase transition, which happens without 
being preceded by the structurally related and simpler B2-phase. The other sharp peak of $C_V$ located at a relatively low temperature of about $200$~K and also complemented by the inflection point of SRO parameters,
is related to the phase separation transition, when the D0$_3$ phase precipitates with the stoichiometric composition inside the pure Fe matrix, tending thereby to the ground state of this alloy at $0$~K.

\begin{figure*}[t!]
    \centering
    \includegraphics[width=0.9\textwidth]{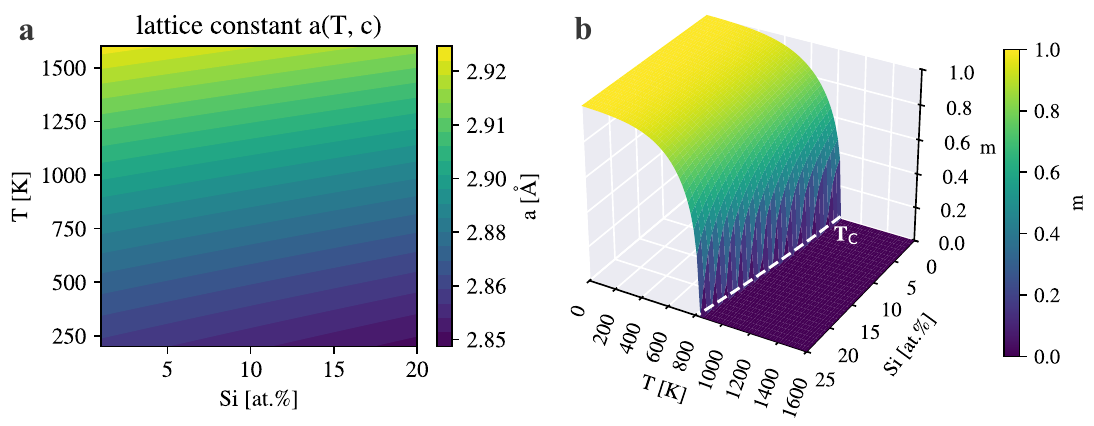}
    \caption{(a) Function for the concentration- and temperature-dependent lattice constant, based on the empirical formula in Ref.~\cite{FeSi_alat_exp} and used for all magnetic states.
    (b) Magnetization dependence on temperature and Si concentration, based on Refs.~\cite{magn_curve_Kuzmin,Fe_alat_exp} and used for PDLM simulations labeled $m(T,c)$.
    }
    \label{fig:grid}
\end{figure*}

The ordering behavior of Fe-Si is quite different in the PM (DLM) state. The high-temperature alloy configuration 
in this case exhibits the B2 type of ordering. The first ordering transition occurs approximately at $800$~K, where B2 
precipitates are formed. This is a primary ordering transition. Secondary ordering occurs at
about $250$~K, where the off-stoichiometric B2 phase transforms to the off-stoichiometric D0$_3$. Finally,
the stoichiometric D0$_3$ phase precipitates in pure Fe at about $50$~K, which is the ground state of this system at $0$~K.

These results clearly show a strong influence of the magnetic state on the ordering behavior
of Fe-rich Fe-Si alloys. Since the magnetic state itself depends on the temperature and composition, this sets the scene for a complex interplay between magnetic and atomic ordering, which will be discussed in the next section.

\section{Results II: Temperature- and concentration-dependent simulations}
\label{ch:ResultsII}

\subsection{Grid setup for SGPM calculations}
\label{ch:grid}

In the previous section, we have seen that both the lattice constant, $a$, and the magnetic state, $m$,
have a strong effect on the interactions and eventually on the ordering behavior.
Since in practice they are controlled by the temperature and the concentration of Si, we
first need to define the functions $a(T, c_{\text{Si}})$ and $m(T, c_{\text{Si}})$.
For the lattice constant, we use an empirical formula derived for Fe-rich Fe-Si alloys in Ref.~\cite{FeSi_alat_exp}, namely
\begin{equation}
    a(T, c_{\text{Si}}) = 2.8658 + 3.747 \cdot 10^{-5} \, T + 8.59 \cdot 10^{-9} \, T^2 - 6.5 \cdot 10^{-2} \cdot c_{\text{Si}},
\end{equation}
where $T$ is the temperature given in Celsius.

The function resulting from this formula can be seen in Fig.~\ref{fig:grid}(a). Note that while this empirical approach is very reliable for the disordered A2 phase, i.e., for the concentrations below approximately 10-15~at\% Si,
it is only approximate for the ordered B2 and D0$_3$ phases. In these phases, the lattice parameter is no longer governed solely by the average chemical composition, but is also significantly influenced by the atomic ordering. Consequently, the empirical relation systematically overestimates the lattice constant. For example, this empirical formula predicts a lattice parameter of approximately $a=2.85$~\AA{} in the ordered D0$_3$-Fe$_3$Si, while the experimentally reported value is around $a=2.83$~\AA. However, for the purposes of the present study, this limitation is not expected to be of significant relevance, as the investigated compositions do not exceed 20~at\%~Si.

\begin{figure*}[t]
    \centering
    \includegraphics[width=0.95\textwidth]{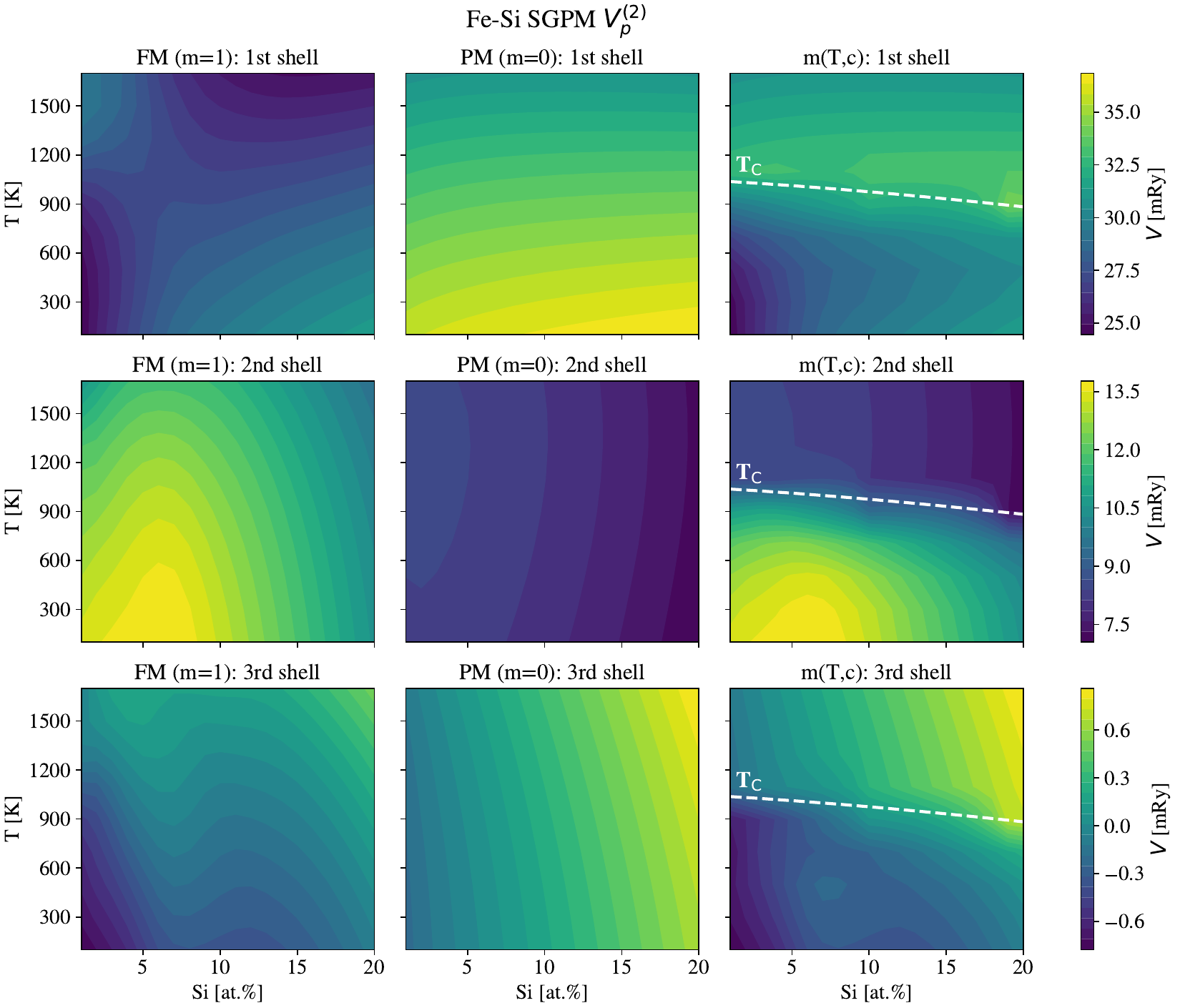}
    \caption{Chemical EPI heat maps at the first three coordination shells obtained by the SGPM for FM ($m=1$), PM ($m=0$), and $m(T,c)$. }
    \label{fig:EPI_maps}
\end{figure*}

Additionally, in order to obtain a realistic picture of the magnetic state close to and above the Curie temperature within PDLM, a function for concentration and temperature dependence of the magnetization $m(T,c_{\text{Si}})$ must be introduced. The starting point here is the function derived by M.~D. Kuz'min \cite{magn_curve_Kuzmin}:
\begin{equation}
m(T) = \left[ 1 - s \left(\frac{T}{T_C} \right)^{3/2} - (1-s) \left(\frac{T}{T_C} \right)^{p} \right]^{1/3}.
\end{equation}
Here, $s$ and $p$ are material-specific shape and exponential parameters, respectively, and $T_C$ is the Curie temperature. We set the exponential parameter to $p=4$, as recommended for bcc Fe, and obtain the shape parameter using the equation
\begin{equation}
    s = \frac{3}{8}\pi^{-3/2} \zeta \left( \frac{3}{2} \right) \frac{\mu_B}{M_0} \left( \frac{k_B T_C}{D} \right)^{3/2},
\end{equation}
where for the zero-temperature spontaneous magnetization $M_0^{\text{Fe}}$ we use the experimental value for Fe $1.76 \cdot 10^6$ A/m, and assume that it decreases linearly with Si concentration as $M_0(c_{\text{Si}}) = (1-1.25 \cdot c_{\text{Si}}) M_0^{\text{Fe}}$. $D$ is the spin-wave stiffness constant which is set to $270$~meV\AA$^{2}$ for the purpose of this work. Evidently, both of the above equations require the Curie temperature, which also strongly depends on solute content in Fe-Si. To obtain the concentration-dependent Curie temperature, we employ a second empirical relation from Ref.~\cite{FeSi_alat_exp},
\begin{equation}
    T_{C}[K] = 1043 \cdot c_{\text{Fe}} + 504 \cdot c_{\text{Fe}} c_{\text{Si}} (c_{\text{Fe}} - c_{\text{Si}}).
\end{equation}
As a result, Si concentration is accounted for in both the Curie temperature $T_{C}$ and the zero-temperature spontaneous magnetization $M_0$. In the end, the shape of the magnetization stays approximately constant for all Si concentrations in the region of interest, with a shape parameter of around $s=0.19$. The final resulting magnetization function is depicted in Fig.~\ref{fig:grid}(b).

\subsection{Concentration and temperature dependence of EPIs}
\label{ch:EPI_grid_results}

\begin{figure*}[t!]
    \centering
    \includegraphics[width=0.98\textwidth]{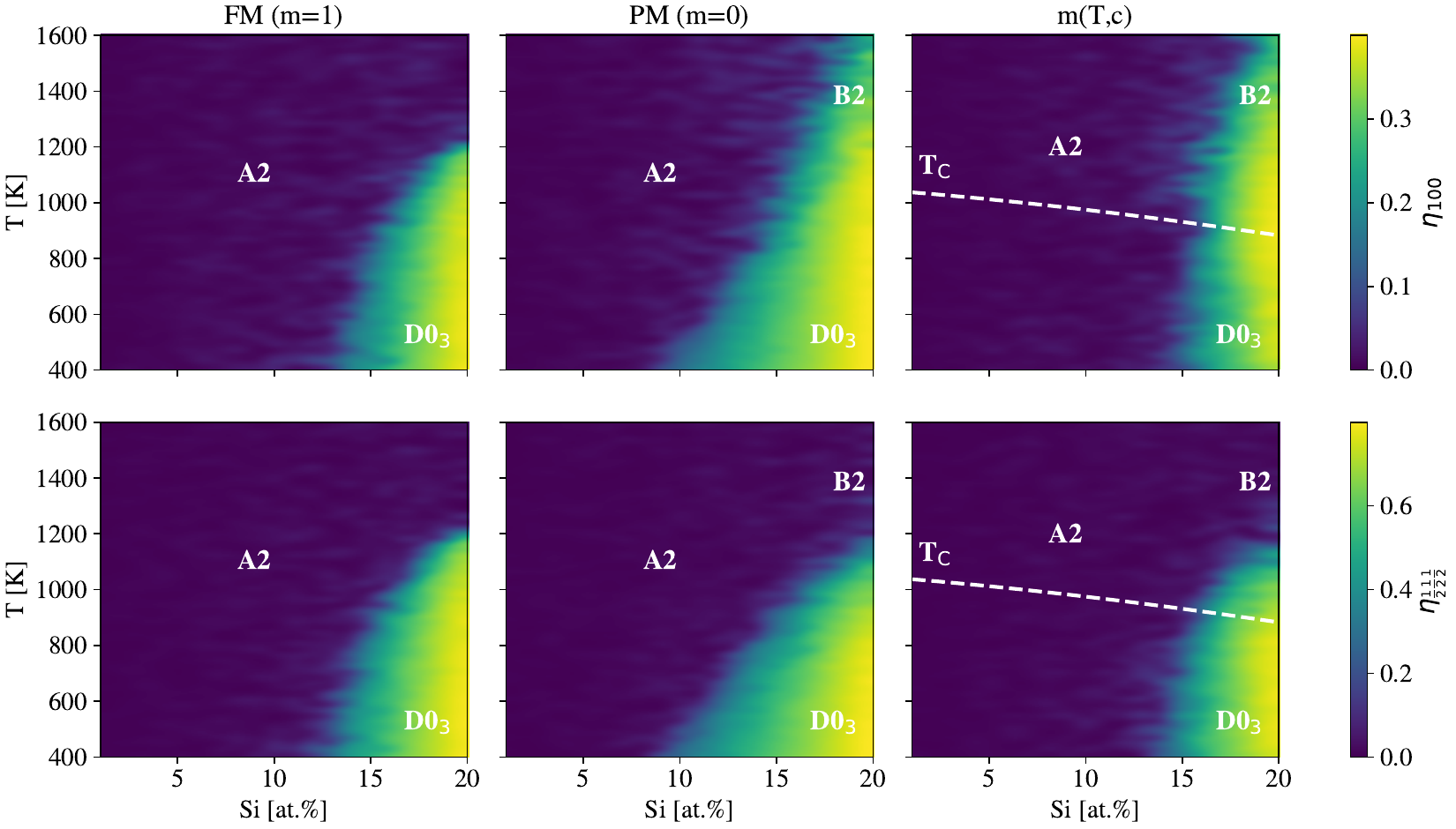}
    \caption{$\eta_{100}$ and $\eta_{\frac{1}{2}\frac{1}{2}\frac{1}{2}}$ LRO parameters obtained in MC simulations for different magnetic states, FM ($m=1$), PM ($m=0$) and $m(T,c)$. The given ordered phases can be identified by
    the values of these parameters: in the D0$_3$ phase both LRO parameters are non-zero, while in the B2 phase only $\eta_{100}$ is non-zero. Notably, the B2 phase is suppressed by FM ordering, even when thermal expansion is accounted for.}  
    \label{fig:cv_lro_maps}
\end{figure*}

Using the functions for lattice constants and magnetization described in Section~\ref{ch:grid}, we perform SGPM calculations for a grid of various lattice constants and Si concentrations in the FM and PM states, where the latter is treated within the DLM approximation. Beyond that, PDLM calculations are performed for $0 < m < 1$, with $m$ taken from the magnetization curve shown in Fig.~\ref{fig:grid}(b). The grid is set up with $0.01$~\AA{} steps in lattice constants and 1~at\% steps in Si content. In this fashion, two-site, three-site, and four-site ECIs are obtained for the entire grid, which are later used in statistical MC simulations. Note that for MC, additional linear interpolation of the SGPM grid is required along the temperature axis, to achieve a desired finer step size of $20$~K.
The results for the first three EPIs are shown in Fig.~\ref{fig:EPI_maps}. For MC simulations in the following Section~\ref{ch:MC_Results}, strain-induced interactions of the first five EPIs, obtained from FM supercell calculations with a lattice constant of $a=2.88$~\AA, are added as constants to the entire grid: $V^{(2)}_{\text{si}} = [-9.95, -5.50, -0.08, -0.18, -0.22]$~mRy. Thus, it is assumed that they do not change significantly with concentration, temperature, and magnetic state. This is, of course, an approximation, but it is practically impossible to obtain the strain-induced interactions for the entire variety of these parameters.

As can be seen in Table~\ref{tab:EPI} and Fig.~\ref{fig:EPI_maps},
the EPIs at the first two coordination shells are large and positive, especially in the FM state. This means that the formation of
Fe-Fe or Si-Si pairs at those coordination shells in the FM state is highly unfavorable. In the DLM state, the interaction at the second coordination shell is significantly reduced compared to that in the FM state. It is further reduced when one adds the strain-induced interaction, so the total EPI at the second coordination shell becomes small and cannot prevent the Fe-Si ordering at the second coordination shell, which makes the stabilization of the B2 phase in the PM state possible.

In the final row in Fig.~\ref{fig:EPI_maps}, we combine the ECIs from FM and DLM together with the PDLM calculations in the regions where the
magnetization is between zero and one. As a result, the final grid for the ECIs should enable MC simulations with a magnetic transition closely related to what one would see in a real system: a FM state at low temperatures, a PM state above the Curie temperature, and partial magnetic ordering
with $0 < m < 1$ when approaching the Curie temperature from below. In Fig.~\ref{fig:EPI_maps}, one can clearly see the region where this magnetic transition occurs.

\subsection{Phase equilibria in Fe-rich Fe-Si alloys from MC simulations}
\label{ch:MC_Results}

\begin{figure*}[t]
    \centering
    \includegraphics[width=0.98\textwidth]{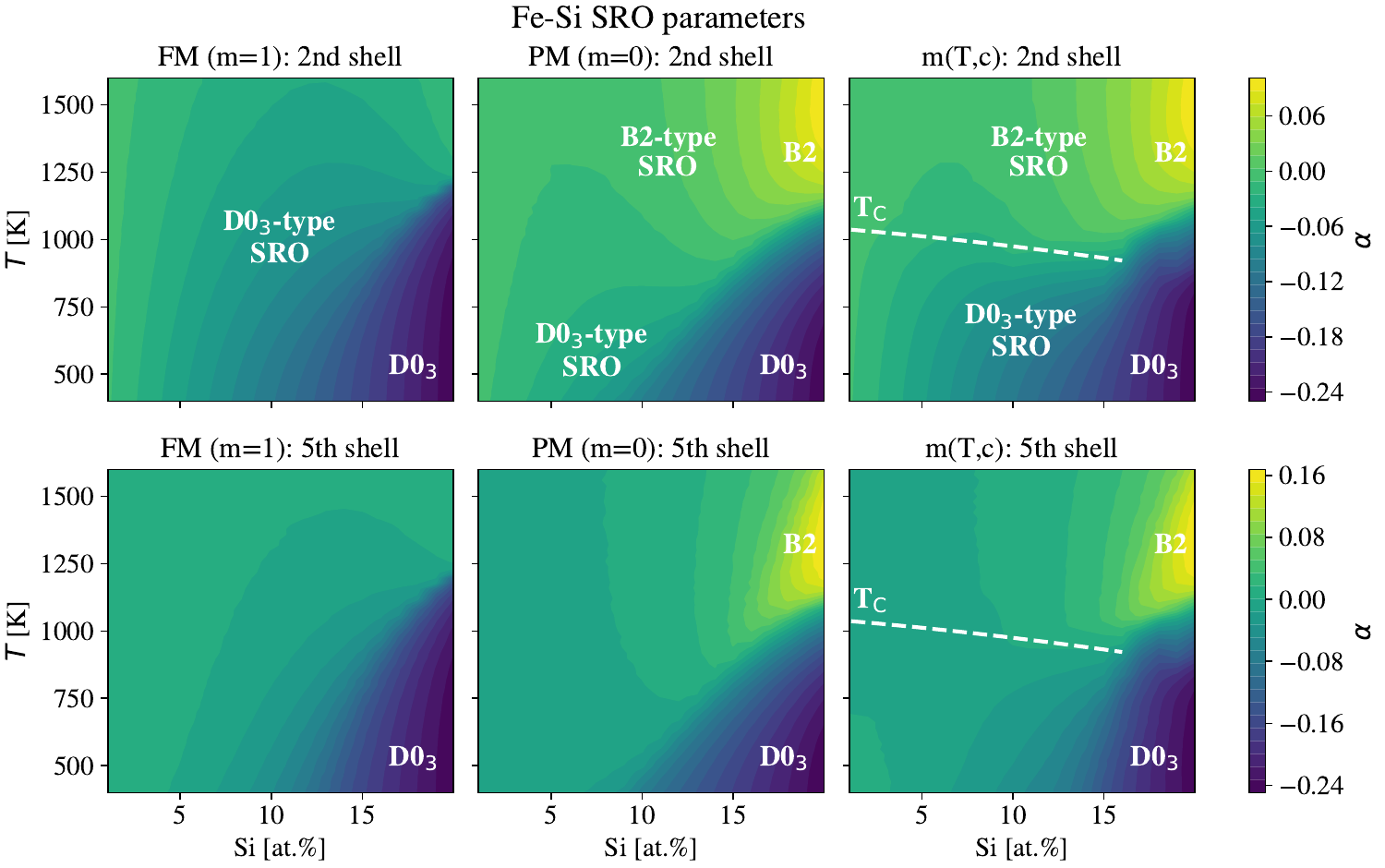}
    \caption{Warren-Cowley SRO parameters for the characteristic second and fifth coordination shells (see Table~\ref{tab:SRO}) for different magnetic states: FM ($m=1$), PM ($m=0$), and $m(T,c)$. SRO parameters in these coordination shells take on positive values in B2 and negative values in D0$_3$. Using this definition, we identify B2- and D0$_3$-type SRO regions of Fe-Si alloys, generally found in the vicinity of the corresponding long-range ordered regions.}
    \label{fig:SRO_parameters}
\end{figure*}

As mentioned above, straightforward calculations of LRO parameters in MC simulations are not always possible. Nevertheless, this can be accomplished approximately by identifying the corresponding Fe and Si sublattices and examining the concentrations on each of them. So, in our MC simulations, the LRO parameters have been obtained by post-processing from the final configurations. 

We have performed MC simulations with the effective interactions on the temperature-concentration grid, as described in the previous section
for the Fe-Si alloy up to 20~at\% Si. The first twenty EPIs have been used in the calculations. As for the higher-order ECIs, we included only the strongest three-site and four-site interactions that affect the ordering energy the most. In particular, we have used up to six three-site and up to four four-site ECIs. To provide the temperature- and concentration-dependent ECIs on a significantly denser temperature grid in the MC simulations than in the SGPM calculations, we have performed linear interpolation of the ECIs for all grid points not calculated directly.
The MC simulation boxes are $12\times12\times12 \, (\times \, 2)$ and the total number of MC steps is 6000, whereas quantities like correlation functions of the atomic distribution, total energies, and heat capacities have been averaged over the last 3000 steps.

In Fig.~\ref{fig:cv_lro_maps}, we present the LRO parameters $\eta_{100}$ and $\eta_{\frac{1}{2}\frac{1}{2}\frac{1}{2}}$ obtained in the MC simulations for various magnetic states FM, DLM, and PDLM. In the latter case, we have used the FM and DLM effective interactions for $m=1$ and $m=0$, respectively. Our simulations show that the formation of the B2 phase is suppressed in the FM state even when the thermal expansion is included.
So, the D0$_3$ phase precipitates directly from the random solid solution in the FM state, as has been seen in Section~\ref{ch:generic_picture}. The B2-phase is stabilized only by magnetic disorder. This stabilization happens in both DLM and PDLM calculations (which is actually equivalent to DLM at temperatures above $T_C$). For an exemplary concentration of Si at around 18~at\%, the first order-disorder transition into the B2-phase in these cases occurs at relatively high temperatures of about $1500$~K, and the second ordering transition to the D0$_3$ structure occurs at about $1000$~K. These results are in fair agreement with the existing experimental data, as will be discussed below.

The Warren-Cowley SRO parameters provide additional information about the type of ordering and can also be used for
identification of ordering phase transitions, as discussed in Section~\ref{ch:generic_picture}. In particular, the SRO parameters
at the second, fifth, and eighth coordination shells are positive in the B2 type of ordering and negative in the D0$_3$ type.
In Fig.~\ref{fig:SRO_parameters}, we show the SRO parameters at the second and fifth coordination shells obtained in MC simulations 
for the FM, DLM, and PDLM states. In the FM state, they are negative, which corresponds to the D0$_3$ type of ordering as discussed
above. In the DLM and PDLM cases, they are positive at high temperatures but become negative below the Curie temperature.

As can be seen for the FM state in Fig.~\ref{fig:SRO_parameters}, the D0$_3$ type of ordering is prevalent across the majority of
the Fe-rich region of Fe-Si alloys, also found beyond the long-range-ordered domain. In the second coordination shell, the D0$_3$-type SRO region extends to low Si concentrations of 5~at\%, where values of around $\alpha_2=-0.05$ to $-0.06$ are obtained from the MC simulations in the low temperature region of $400$~K to $600$~K, a clear deviation from a fully random structure. At the same time, in the PM ($m=0$) state, large sections of the Fe-rich region are restricted to the B2 type of ordering, and only a small region at very low temperatures aligns with the definition of D0$_3$-type SRO. Finally, for the SRO parameters of the second shell of the $m(T,c)$ result, the D0$_3$-type SRO region is extended relative to the DLM result, due to the FM ordering below the Curie temperature.

The SRO parameters of the fifth coordination shell, depicted in  Fig.~\ref{fig:SRO_parameters}, can be treated as an alternative quantification of LRO.
In the ordered regions, the SRO parameters are found to be in agreement with the results from LRO parameters previously shown in Fig.~\ref{fig:cv_lro_maps} and Table~\ref{tab:SRO}. In particular, off-stoichiometric B2 long-range ordering is suppressed in the FM state, and the corresponding SRO parameters in the second and fifth shells take on only negligibly small positive values in the simulated region.

\begin{figure*}[t]
    \centering
    \includegraphics[width=\textwidth]{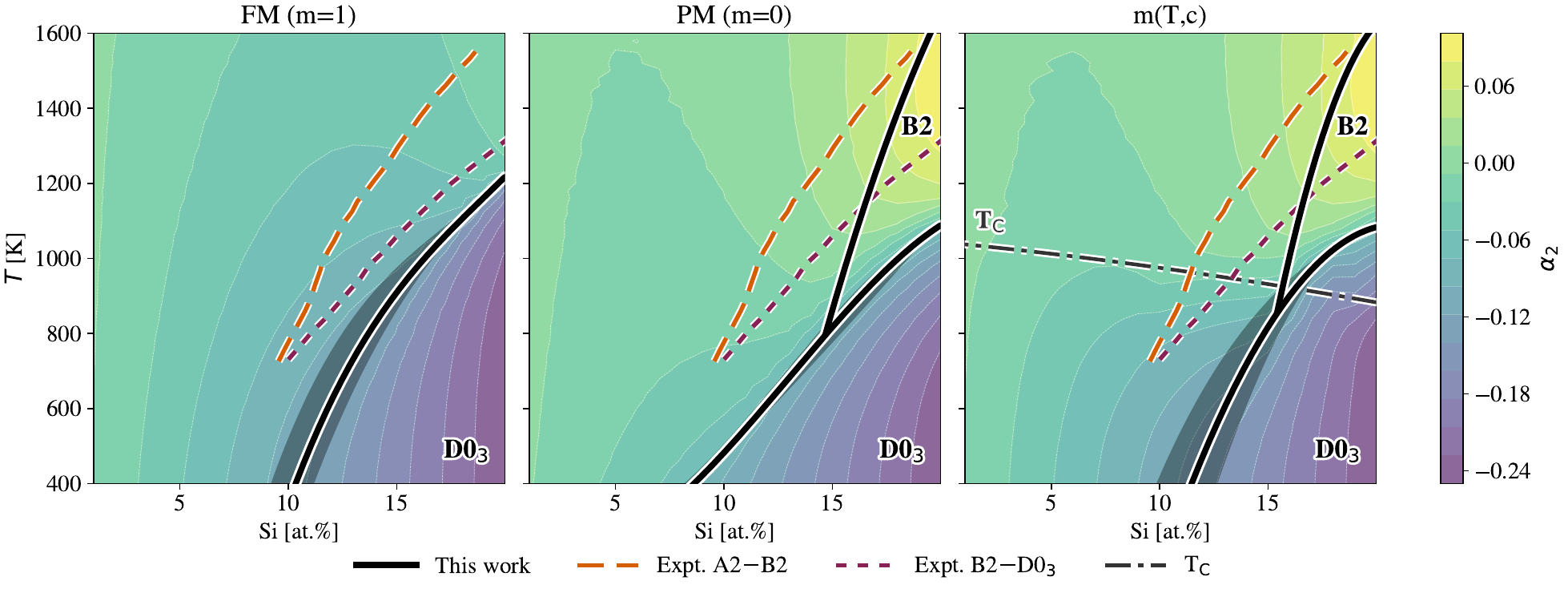}
    \caption{Calculated Fe-Si phase diagrams in FM ($m=1$), PM ($m=0$) and $m(T,c)$ magnetic states. Solid lines refer to MC simulations and are fitted using the SRO parameters of the characteristic eighth coordination shell (see Table~\ref{tab:SRO}). Experimental phase boundaries are shown with dashed lines (obtained from Ref.~\cite{FeSi_calphad}). The background contour plots show the SRO parameters of the second coordination shell, $\alpha_2$, which take on positive values for B2 and negative values for D0$_3$. Note that among the MC results shown here, $m(T,c)$ is the most accurate approximation for the magnetic ordering of the experimental data~\cite{FeSi_calphad}, whereas the FM and PM results refer to the low- and high-temperature limits, respectively.}
    \label{fig:phasediag}
\end{figure*}

Finally, we use a similar idea to draw an estimate of the phase diagram, as seen in Fig.~\ref{fig:phasediag}. Here, the SRO parameters
at the eighth coordination shell are used as an indicator of the type of LRO. This shell, as well as the second and fifth shells, take on values of $-\frac{c}{1-c}$ for the D0$_3$ phase, and values of $\frac{c}{1-c}$ for the perfect off-stoichiometric B2 (see Table~\ref{tab:SRO}), with no sub-ordering in the Si-sublattice and no precipitation. Considering this, the transition lines are fitted through the SRO parameters normalized by a factor of $\frac{1-c}{c}$ to remove the dependency on the concentration and to map the parameters into an interval $[-1,1]$. The fit, then, passes through $0.3$ for the A2-B2 transition line, and $-0.3$ for the A2-D0$_3$ transition line. The B2-D0$_3$ line passes through the points where the SRO parameters are zero. The shaded area around the B2-D0$_3$ line is drawn for the normalized SRO parameters between $-0.2$ and $-0.4$. As one can see this area is particularly large in the FM region.
In these simulations, SRO parameters are averaged over $3000$ MC configurations. Considering this, any deviations from zero in the SRO parameter $\alpha_8$ at the eighth coordination shell indicate that at least some configurations in this set of structures contain LRO. In the FM region of the phase diagram, $\alpha_8$ tends more slowly to zero than in the case of DLM, suggesting that a subset of the sampled configurations appear as long-range ordered.
As a result, the exact position of this transition line is hard to identify, and it can therefore be interpreted only approximately. Note that a more accurate description of the order-disorder lines would require the use of the grand canonical ensemble and evaluation of Gibbs free energies, which is beyond the scope of this study.

\begin{figure*}[t]
    \raggedright
    \includegraphics[width=0.94\textwidth]{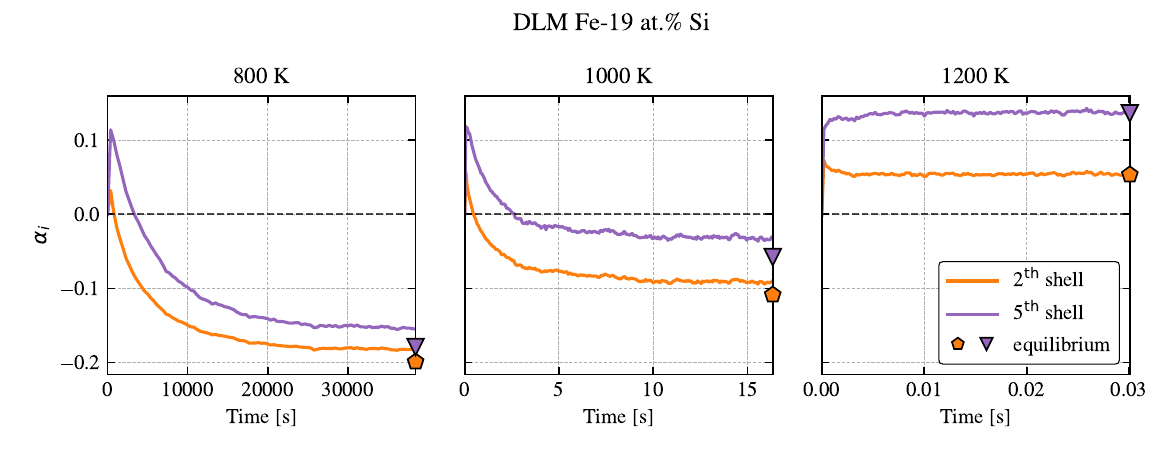}
    \caption{Warren-Cowley SRO parameters obtained from Kawasaki dynamics kMC, as a function of estimated physical time, for PM (DLM) Fe–Si with 19~at\%~Si at temperatures $800$~K, $1000$~K, and $1200$~K. The characteristic SRO parameters $\alpha_2$ and $\alpha_5$ are initially positive for all temperatures, indicating the initial formation of B2-type ordering, regardless of the final ordered equilibrium phase.}
    \label{fig:kMC_result}
\end{figure*}

\subsection{Dynamical properties of chemical ordering in Fe-Si alloys}
\label{ch:kMC_results}

The MC results presented above describe the phase equilibria in Fe-Si alloys. However, for practical applications, 
the time evolution of chemical ordering is also of considerable interest. In this work, we use kinetic MC simulations based on
Kawasaki dynamics, as described in Section~\ref{ch:MC}, to estimate the time required for the atomic ordering to reach equilibrium, starting from perfectly disordered initial configurations.
The simulated MC steps are converted to physical time using the Arrhenius relation in Eq.~\ref{eq:Arrh_law}. Here, we use the pre-exponential factor $\tau_0=1.7\times10^{-19}$s and the activation energy $Q=69 \times 10^{3}$ cal/mol, both taken from Ref.~\cite{Diff_FeSi}. Additionally, the authors of that paper also discuss a geometry factor $\gamma$, unique to their experimental setup, required for obtaining correct diffusion times. In the following, we use the suggested factor of $\gamma=1/16$. It should be emphasized that these parameters carry large experimental errors, and are measured for a lower Si content of 4~wt\% (approx.~8~at\%). Thus, our results should be taken as a rough estimate, especially taking into consideration the fact that diffusion in alloys with large absolute values of SRO parameters, like the ones we have obtained at higher Si contents, can be significantly slower.

In Fig.~\ref{fig:kMC_result}, we show the time evolution of the Warren-Cowley SRO parameters of the second and fifth coordination shells in the Fe–19~at\% Si alloy in the PM (DLM) state for three temperatures: $800$~K, $1000$~K, and $1200$~K, spanning the region of the B2 and D0$_3$ phase formation
(see Fig.~\ref{fig:SRO_parameters}). Additionally, the equilibrium values of these SRO parameters obtained from the conventional MC, as described in Section~\ref{ch:MC_Results}, have been added to the right side of each figure. As in the previous sections, the SRO parameters allow one to distinguish between the B2-type and D0$_3$-type ordering. At all three temperatures, the first few nearest-neighbor exchanges immediately produce positive SRO parameters, indicating the rapid formation of the
B2 phase. However, the subsequent time evolution strongly depends on the specific temperature.

At $1200$~K, all SRO parameters remain positive throughout the simulation, quickly reaching the thermodynamically stable B2 phase also obtained in the conventional MC simulation. At $1000$~K, the SRO parameters gradually decrease, with the fifth shell crossing over into negative values after only a few seconds, indicating a gradual transition from the B2 phase to the D0$_3$ type of ordering. Finally, at $800$~K, the same ordering reversal occurs, but on a significantly extended timescale. In the last two cases, at $1000$~K and $800$~K,
the D0$_3$ phase is in fact the equilibrium one, as can be seen from the equilibrium values in Fig.~\ref{fig:kMC_result}, or Fig.~\ref{fig:SRO_parameters} from the previous section. This means that the initial formation of the
B2 type of ordering is related to atomic correlations from the exceedingly strong interaction at the first coordination shell, and
the geometry of the lattice, but this state is far from equilibrium for $T \leq 1000$~K.

\section{Conclusions}
\label{ch:Conclusion}

The present \textit{ab initio} simulations demonstrate that an accurate description of finite-temperature magnetism is essential for predicting chemical ordering in Fe–Si alloys.
To account for these effects, we employed effective chemical interactions obtained from SGPM and complementary VASP supercell calculations. These interactions exhibit a strong dependence on magnetic state, composition, lattice parameter, and exchange-correlation functional. When these dependencies are properly incorporated, the resulting model successfully reproduces the experimentally observed ordering behavior of Fe-rich Fe-Si alloys.

The calculations reveal a strong interplay between magnetism and atomic ordering. The B2 phase is stabilized only in the paramagnetic state and appears exclusively as an off-stoichiometric ordered phase in Fe-rich alloys. In contrast, ferromagnetic order promotes the subsequent ordering transition toward the D0$_3$ phase, which is  off-stoichiometric under experimentally relevant conditions. Magnetism also strongly influences atomic short-range order (SRO) in the disordered A2 phase. While ferromagnetic alloys exhibit only D0$_3$-type SRO, both B2- and D0$_3$-type correlations coexist in the paramagnetic state.

The time evolution of the Warren-Cowley SRO parameters was further investigated by kinetic Monte Carlo simulations in the vicinity of the B2-D0$_3$ transition. The results show that the B2 phase forms immediately in the simulations based on the initially random configurations, independent of temperature and regardless of whether the equilibrium state is ultimately B2 or D0$_3$. This behavior reflects the dominant influence of the strong nearest-neighbor effective interactions, which drive the ordering process and subsequently affect correlations at more distant coordination shells.

Although the predicted phase equilibria agree well with experiment up to 20~at\% of Si, the description of SRO in alloys containing less than 10~at\% Si remains largely semi-quantitative. In this regime, pronounced local-environment effects arise from the influence of Si atoms on the magnetic moments of neighboring Fe atoms, even in the paramagnetic state. As a consequence, the assumptions underlying the present Ising-like description become increasingly restrictive, indicating that a more advanced \textit{ab initio} description with an explicit treatment of local magnetic fluctuations will be required for fully quantitative predictions. Such extensions will be the subject of future work.

\acknowledgements
The financial support by the Austrian Federal Ministry of Economy, Energy and Tourism, the National Foundation for Research, Technology and Development and the Christian Doppler Research Association is gratefully acknowledged. 
We gratefully acknowledge the computational resources provided by the Vienna Scientific Cluster.
A significant part of the computations was enabled by resources provided by the National Academic
Infrastructure for Supercomputing in Sweden (NAISS), partially funded by the Swedish Research
Council through grant agreement no. 2022-06725. Support of
the staff of the National Supercomputing Center (Linköping) and PDC Center for High Performance Computing
at KTH is appreciated. We acknowledge access to LEONARDO at CINECA, Italy, via an 
AURELEO (Austrian Users at LEONARDO supercomputer) project.
This research was funded in whole or in part by the Austrian Science Fund (FWF) [10.55776/I6756]. The financial support by voestalpine Stahl GmbH is gratefully acknowledged.
\newpage

\appendix

\section{Higher-order ECI from grid-based SGPM}

Three-site and four-site ECI where found to be crucial for D0$_3$ ordering in Fe-Si alloys. When they are neglected entirely, this phase often does not precipitate at low temperatures, particularly for the FM state, where higher-order ECI from SGPM take on large values commonly found in the range of $2$~mRy to $4$~mRy. In contrast, in DLM we found higher-order ECI to be less significant.

Higher-order ECI are commonly defined using indices specified by the coordination shells.~\cite{ECI_fcc} For example, if the distances between atoms in a given three-site cluster are all equal to the first coordination shell, the cluster would be indexed as $111$. If two distances are the first coordination shell, and one is the second it would be $112$, and so on. One might notice that not every cluster can be constructed in every crystal system, e.g., the $111$ cluster exists in fcc, but does not in bcc or sc.

The same logic is used for four-site interactions. Here, there is a total of six connections between the four atoms of each cluster, so six indices are required to uniquely define them.

In this study of bcc Fe-Si alloys the relevant ECI that were also used across all MC simulations are $112$, $113$, $223$, and $115$ for three-site clusters, and $112211$, $111223$, and $111333$ for four-site clusters. Note that no strain-induced interactions were added to three-site and four-site interactions, which could also impact ordering effects in these systems.

For FM SGPM many more ECI were found to have large absolute values. For example, when searching the full grid for ECI with absolute values larger than $2$~mRy, the following are obtained in addition to the above: $124$, $226$, $149$, $112311$, $112214$ and $112514$. However, adding them to MC simulations was found to often be detrimental for D0$_3$ ordering. We assume that missing strain-induced properties, vibrational effects, magnetic local-environment effects, or the counterplay with other higher-order ECI are likely the reasons for that. The computational cost of accounting for strain-induced properties of such a high number of ECI was beyond the scope of this study. Note that for DLM SGPM absolute values of higher-order ECI are significantly smaller, and with the same threshold of $2$~mRy no additional ECI are found to be relevant for the entire grid over lattice constants and Si concentrations.

\newpage
\bibliographystyle{apsrev4-2}
\bibliography{refs}

\end{document}